\documentclass[11pt]{article}
\usepackage{moriond,epsfig}
\usepackage{amsmath}

\def\be{\begin{equation}}
\def\ee{\end{equation}}
\def\bea{\begin{eqnarray}}
\def\eea{\end{eqnarray}}

\begin{document}
\vspace*{4cm}
\title{Implications of new generations on neutrino masses}

\author{A. APARICI, J. HERRERO-GARCIA, N. RIUS and A. SANTAMARIA}

\address{Depto.de Fisica Teorica, and IFIC, Universidad de Valencia-CSIC \\ Edificio de Institutos de Paterna, Apt. 22085, 46071 Valencia, Spain}

\maketitle\abstracts{
We explore the possible implications that new families, that are being searched for at the LHC, would have on neutrino masses. In particular, we have explored the possibility that the smallness of the observed neutrino masses is naturally understood in a modified version of the Standard Model (SM) with complete extra generations of fermions, i.e., that have right-handed neutrinos, in which neutrino masses are generated at two loops. With one extra family it is not possible to fit the observed spectrum of masses and mixings. However, the radiative mass generated provides an important constraint in these kind of models, so the neutrino masses do not exceed their cosmological bound. Within the context of two extra families, we analyse the allowed parameter space and the possible phenomenological signals. Contribution to NUFACT 11, XIIIth International Workshop on Neutrino Factories, Super beams and Beta beams, 1-6 August 2011, CERN and University of Geneva (Submitted to IOP conference series).}

\section{Introduction}
One of the most natural extensions of the Standard Model (SM) is the addition of extra sequential generations \cite{Holdom:2009rf}. We analyse the relation between new generations and neutrino masses, one of the strongest evidence for physics beyond the SM, in view of the possibility that the former are discovered at the LHC.

Global fits of models with additional generations to the electroweak data have been performed and they favour no more than five generations. It should be kept in mind that most of the fits make some simplifying assumptions on the mass spectrum of the new generations and do not consider Majorana neutrino masses for the new generations or the possibility of breaking dynamically the gauge symmetry via the condensation of the new generations' fermions; all these would give additional contributions.

LEP II limits on new generation leptons are: $m_{\ell'}>100.8$ GeV and $m_{\nu'}>$ $80.5\:(90.3)\:\mathrm{GeV}$ for pure Majorana (Dirac) particles. When neutrinos have both Dirac and Majorana masses, the bound on the lightest is $63\,\mathrm{GeV}$. For stable neutrinos LEP I measurement of the invisible $Z$ width, $\Gamma_{\mathrm{inv}}$, implies $m_{\nu'}>39.5\,(45)\,\mathrm{GeV}$ for pure Majorana (Dirac) particles. Recent LHC bounds on the quarks are $m_{t'}>450$ GeV, assuming $t' \rightarrow b W$, which in principle is plausible as EW fits bound the splitting within the doublets to be $|m_{t'}-m_{b'}| < 80$ GeV, and $m_{b'}>385$ GeV. For the leptons, the splitting can be larger, $|m_{\ell'}-m_{\nu'}| < 140$ GeV.

The most striking effect of new generations is the enhancement of the Higgs-gluon-gluon vertex which arises from a triangle diagram with all quarks running in the loop, by a factor of 9 (25) in the presence of a fourth (fifth) generation. Thanks to that, LHC  has been able to exclude, at $95 \%$ C.L., a fourth-generation SM-like Higgs in the range $120<m_{H}<600\,\mathrm{GeV}$. Stability of the Higgs potential, and triviality, together with these severe bounds, imply that if new generations exist, they will presumably come together with a new non-standard scalar sector, for instance, a strongly coupled composite Higgs, an extended Higgs sector or a Higgs with invisible decays, for example, to dark matter. Of course, if both a new family and the Higgs with a 3 generations' Standard Model cross section are detected, new physics would be needed to reconcile these facts.

In this talk (see \cite{Aparici:2011nu} for further details and a complete list of references) we focus on how neutrino masses can be naturally generated at two loops by adding extra families and singlets. Recall that right-handed neutrinos do not have gauge charges and are not needed to cancel anomalies, therefore their number is not linked to the number of generations.

\section{Four generations\label{SM4}}

We extend the SM by adding a complete fourth generation, i.e., with one right-handed neutrino $\nu_{R}$, with a Majorana mass term \cite{Petcov:1984nz}. We denote the new charged lepton $E$ and the new neutrino $\nu_{E}$. $y_{i}$ with $i=e,\mu,\tau,E$ are the Yukawas of the doublets with $\nu_{R}$. After spontaneous symmetry breaking (SSB) the mass matrix for the neutral leptons at tree level is a $5\times5$ Majorana symmetric matrix which has only one right-handed neutrino Majorana mass term. Therefore, it leads to  three massless Weyl neutrinos and two massive Majorana $\nu_{4}$ and $\nu_{\bar{4}}$, of masses $m_{4,\bar{4}}=\frac{1}{2}\left(\sqrt{m_{R}{}^{2}+4m_{D}^{2}}\mp m_{R}\right)$, where $m_{D}/v=\sqrt{\sum_{i}y_{i}^{2}}\approx y_{E}$, with $v=\langle\phi^{(0)}\rangle$, and $\tan^{2}\theta=m_{4}/m_{\bar{4}}$. Phenomenologically $m_{R}<20$ TeV, so $m_{4,\bar{4}}>63$ GeV. If $m_{R}\ll m_{D}$, we have $m_{4}\approx m_{\bar{4}}$ and $\tan\theta\approx1$ (pseudo-Dirac limit) while when $m_{R}\gg m_{D}$, $m_{4}\approx m_{D}^{2}/m_{R}$, $m_{\bar{4}}\approx m_{R}$ and $\tan\theta\approx m_{D}/m_{R}$ (see-saw limit).

\begin{figure}
\centering{}
\psfig{figure=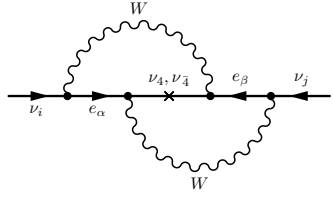,height=1in}
\caption{Two-loop diagram contributing to neutrino masses in the four-generation model. \label{fig:TwoW}}
\end{figure}
Since lepton number is broken by the $\nu_{R}$ Majorana mass, there is no symmetry which prevents the tree-level massless neutrinos from gaining Majorana masses; they are generated at two loops by the diagram of figure~\ref{fig:TwoW}. The eigenvalues of the light neutrino mass matrix are proportional to $m_{\mu}^{4},\, m_{\tau}^{4},\, m_{E}^{4}$ which gives a huge hierarchy between neutrino masses:
\begin{equation}
\frac{m_{2}}{m_{3}}\le\frac{1}{4y_{E}^{2}}\left(\frac{m_{\tau}}{m_{E}}\right)^{2}\left(\frac{m_{\tau}}{m_{\bar{4}}}\right)^{2}\le\frac{10^{-7}}{y_{E}^{2}}\ ,\label{eq:ratiomasasligerosapprox}\end{equation}
where we have used that $m_{E},m_{\bar{4}}\ge100\,\mathrm{GeV}$.
To overcome this huge hierarchy very small values of $y_{E}$ are needed, which would imply that the heavy neutrinos are not mainly $\nu_{E}$ but some combination of the three known ones $\nu_{e},\nu_{\mu},\nu_{\tau}$; however this is not possible since it would yield observable effects in a variety of processes, like $\pi\rightarrow\mu\nu$, $\pi\rightarrow e\nu$, $\tau\rightarrow e\nu\nu$... These require that $y_{e,\mu,\tau}<10^{-2}y_{E}$, so $y_{E}\approx1$. Therefore, the simplest version of the model is unable to accommodate the observed spectrum of neutrino masses and mixings. 

\begin{figure}
\centering{}
\psfig{figure=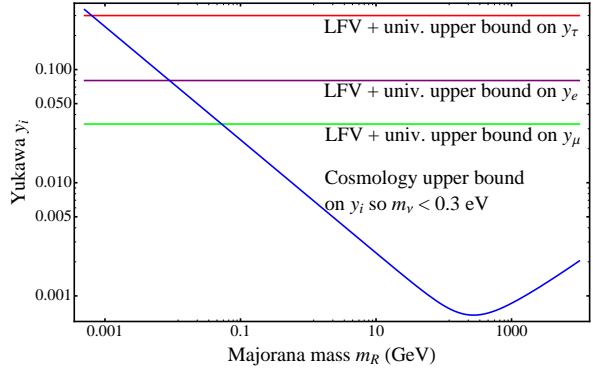,height=2in}
\caption{General constraint on light neutrino masses in four generations' models with heavy Majorana neutrinos. We show the bounds that LFV and universality from pion and tau decays set on the Yukawas at $90 \%$ C.L.: $y_{\tau}<0.3$, $y_{e}<0.08$ and $y_{\mu}<0.033$. \label{fig:cosmo}}
\end{figure}

However, notice that whenever a new generation and a right-handed neutrino with Majorana mass at (or below) the TeV scale are added to the SM, the two-loop contribution to neutrino masses is always present and provides an important constraint for this kind of SM extensions. In figure~\ref{fig:cosmo} it is shown the upper bound in the Yukawa ($y_{i}$, with $i=e, \mu, \tau$) versus Majorana mass plane, so $m_{\nu}<0.3$ eV, as implied by cosmology. It is clearly seen that, for most of the parameter space ($m_{R}>0.01$ GeV), it is more restrictive than universality or LFV constraints.

It could be argued that new generation neutrinos may be Dirac particles, and the bound would not apply. However, if light neutrinos turn out to be Majorana, it is natural that the heavy ones are Majorana as well, since lepton number is not conserved anymore. For instance, in a four generations context with light neutrino masses 
obtained via the see-saw mechanism, as long as it is not completely decoupled from the light neutrinos, the fourth right-handed neutrino will always acquire a Majorana mass at the two-loop level, even if it is set to zero by hand at the Lagrangian level (in particular at the TeV scale for typical GUT scale Majorana masses of the first three right-handed neutrinos), so the previous constraint will in general apply to the light neutrinos' Yukawas.

\section{The five-generation model\label{SM5}}

We add two families to the SM and two right-handed neutrinos, $\nu_{4R}$ and $\nu_{5R}$. We denote the two charged leptons by $E$ and $F$. The model, contrary to the four-generation case, has additional sources of CP violation in the leptonic sector, however, for simplicity we take all neutrino Yukawas $y_{\alpha}$ and $y_{\alpha}^{\prime}$ real. For simplicity, we also choose $\sum_{\alpha}y_{\alpha}y_{\alpha}^{\prime}=0$. 

The tree-level spectrum contains three massless neutrinos and four heavy Majorana neutrinos. The model should be compatible with the observed universality of fermion couplings and have small rates of lepton flavour violation (LFV) in the charged sector, which requires $y_{e},y_{\mu},y_{\tau},y_{e}^{\prime},y_{\mu}^{\prime},y_{\tau}^{\prime} \ll y_{E},y_{F},y_{E}^{\prime},y_{F}^{\prime}$. In addition, it should fit the observed pattern of masses and mixings, for instance, reproducing the tribimaximal (TBM) mixing structure. A successful choice of the Yukawas to obtain normal hierarchy (NH) (see  \cite{Aparici:2011nu} for an analysis of inverted hierarchy (IH) and more details), i.e, $m_{3}\approx\sqrt{\left|\bigtriangleup m_{31}^{2}\right|}\approx0.05$~eV,
$m_{2}\approx\sqrt{\bigtriangleup m_{21}^{2}}\approx0.01$~eV, will be $y_{\ensuremath{\alpha}}=y_{E}(\epsilon,\epsilon,-\epsilon,1,0)\nonumber$ and $y_{\alpha}^{\prime}=y_{F}^{\prime}(0,\epsilon^{\prime},\epsilon^{\prime},0,1)$. Assuming that $m_{E,F}\gg m_{4,\bar{4},5,\bar{5}}\gg m_{W}$, we find: 
\begin{equation}
m_{2}\approx\frac{3g^{4}}{2(4\pi)^{4}m_{W}^{4}}\epsilon^{2}m_{4D}^{2}m_{4R}m_{E}^{2}\ln\frac{m_{E}}{m_{\bar{4}}}\label{eq:m2-NH}\end{equation}
 \begin{equation}
m_{3}\approx\frac{g^{4}}{(4\pi)^{4}m_{W}^{4}}\epsilon^{\prime2}m_{5D}^{2}m_{5R}m_{F}^{2}\ln\frac{m_{F}}{m_{\bar{5}}}\ ,\label{eq:m3-NH}\end{equation}
and the required ratio $m_{3}/m_{2}\approx5$ can be easily accommodated.

\begin{figure}
\psfig{figure=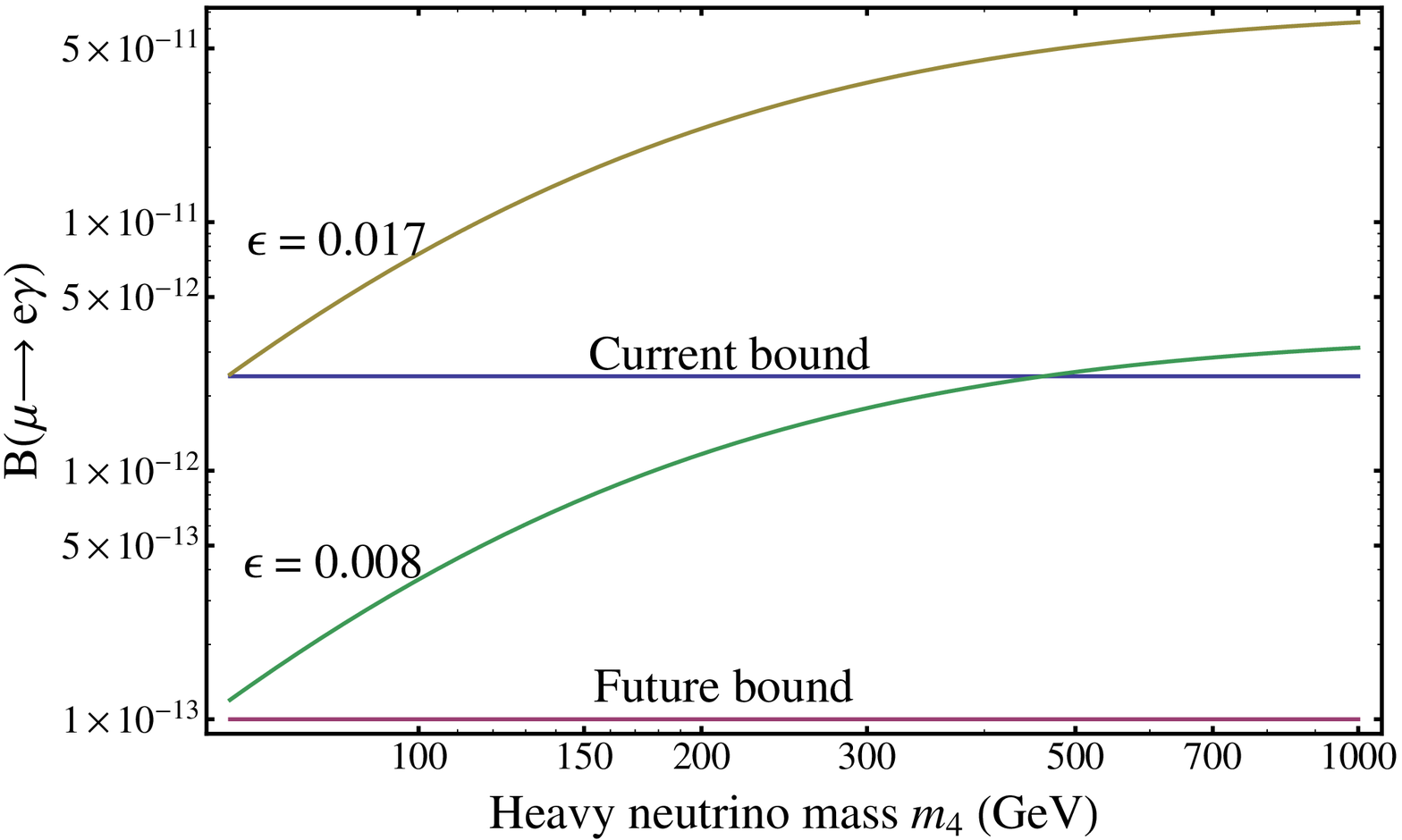, height=1.8in}
\qquad
\psfig{figure=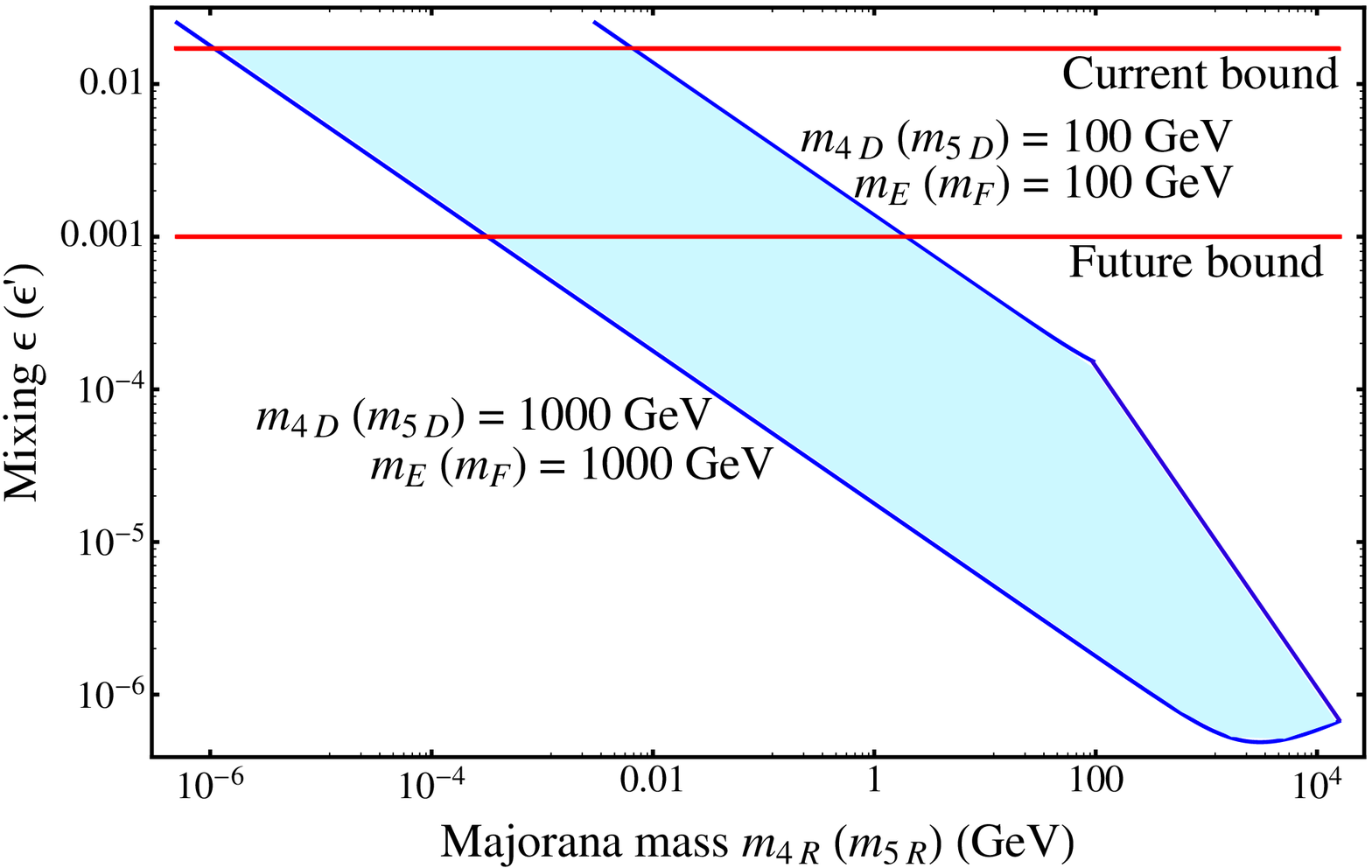, height=1.8in}
\caption{a) Left: $B(\mu\rightarrow e\gamma)$ against $m_{4}$ for different values of $\epsilon$. We also display present and future limits. \label{fig:mu2egamma}
b) Right: Parameter space that predicts the right scale for heavy and light neutrinos (the region between the curves). We also present the current $\mu\rightarrow e\gamma$ bound and the expected $\mu$--$e$ conversion limit. \label{fig:Parameter-space}}
\end{figure}

The new upper bound on LFV $B(\mu\rightarrow e\gamma)<2.4\times10^{-12}$ is translated into $\epsilon<0.02$. We display in figure~\ref{fig:mu2egamma} a) $B(\mu\rightarrow e\gamma)$ versus the mass of the heavy neutrinos $m_{4}$ in the NH case. We also display present and near future limits. Violations of universality constrain the model in both hierarchies. For example, from pion decay, we obtain $\epsilon'<0.04$. 

Regarding lepton number violation signals at the LHC, for instance, if the lightest fourth generation lepton is the charged one, there is a striking signal: lepton number violating like-sign fourth generation
lepton pair-production, through $q\bar{q}'\rightarrow W^{\pm}\rightarrow E^{\pm}\nu_{4,\bar{4}}\rightarrow E^{\pm}E^{\pm}W^{\mp}$ or via $W$ fusion, $q\bar{q}\rightarrow W^{\pm}W^{\pm}q'q'\rightarrow E^{\pm}E^{\pm}jj$.
Also the Higgs could be detected in this way: $gg \rightarrow H \rightarrow \nu_{4}\nu_{4} \rightarrow E^{\pm}E^{\pm}W^{\mp}W^{\mp}$, giving a rather unusual signal, with hardly any background.

With respect to neutrinoless double beta decay, there are contributions of the new heavy neutrinos, however, we obtain the same combination of parameters as in the light neutrino masses expressions, $m_{4R}\epsilon^{2}$, leading to unobservable effects in $0\nu\beta\beta$ when the former are fitted (of course the light neutrino contribution could be observed in the future, if the IH spectrum is realized). 

To summarize the phenomenology of the model we present in figure~\ref{fig:Parameter-space} b) the allowed regions in the $\epsilon-m_{4R}$ plane which lead to $m_{3}\sim0.05\,\mathrm{eV}$ varying the charged lepton masses $m_{E}\,(m_{F})$ and the Dirac neutrino masses $m_{4D}\,(m_{5D})$ between 100~GeV--1~TeV, and imposing the bound on the neutrino mass, $m_{4}>63\,\mathrm{GeV}$. We also plot the present bounds on the mixings $\epsilon\,(\epsilon^{\prime})$ from $\mu\rightarrow e\gamma$ and future much stronger limits from $\mu$--$e$ conversion if expectations are attained. So with five generations and two singlets all current data can be accomodated in the shaded region of the parameter space between the curves of figure~\ref{fig:Parameter-space} b), which will be probed in the near future.

\section*{Acknowledgments}
This work has been partially supported by the grants FPA-2007-60323, FPA-2008-03373, CSD2007-00060, CSD2007- 00042, PROMETEO/2009/116, PROMETEO/2009/128 and MRTN-CT-2006-035482 (FLAVIAnet). A.A. and J.H.-G. are supported by the MICINN under the FPU program.

\section*{References}
\providecommand{\href}[2]{#2}\begingroup\raggedright\endgroup


\end{document}